\documentclass[
reprint,
superscriptaddress,
amsmath,
amssymb,
aps,
longbibliography,
pra, 
showpacs,
floatfix
]{revtex4-1}

\usepackage{graphicx}
\usepackage{color}

\usepackage{tabularx}
\usepackage{dcolumn} 
\newcolumntype{d}[1]{D{.}{.}{#1}}
\usepackage{booktabs}

\usepackage{enumitem}
\usepackage{comment}
\newcommand{\ang}{\ensuremath{\text{\AA}}}

\begin{document}

\title{Orbital magnetic moments of phonons}

\author{Dominik~M.\ Juraschek}
\affiliation{Materials Theory, ETH Zurich, CH-8093 Z\"{u}rich, Switzerland}
\author{Nicola~A.\ Spaldin}
\affiliation{Materials Theory, ETH Zurich, CH-8093 Z\"{u}rich, Switzerland}

\date{\today}


\begin{abstract}

In ionic materials, circularly polarized phonons carry orbital magnetic moments that arise from circular motions of the ions, and which interact with other magnetic moments or fields. Here, we calculate the orbital magnetic moments of phonons in 35 different materials using density functional theory, and we identify the factors that lead to, and materials that show, large responses. We compute the resulting macroscopic orbital magnetic moments that can be induced by the excitation of coherent phonons using mid-infrared laser pulses, and we evaluate the magnitudes of the phonon Zeeman effect in a strong magnetic field. Finally, we apply our formalism to chiral phonons, in which the motions of the ions are intrinsically circular. The zoology presented here may serve as a guide to finding materials for phonon and spin-phonon driven phenomena.
\end{abstract}

\maketitle


\section{Introduction}

\begin{figure}[t]
\includegraphics[scale=0.07]{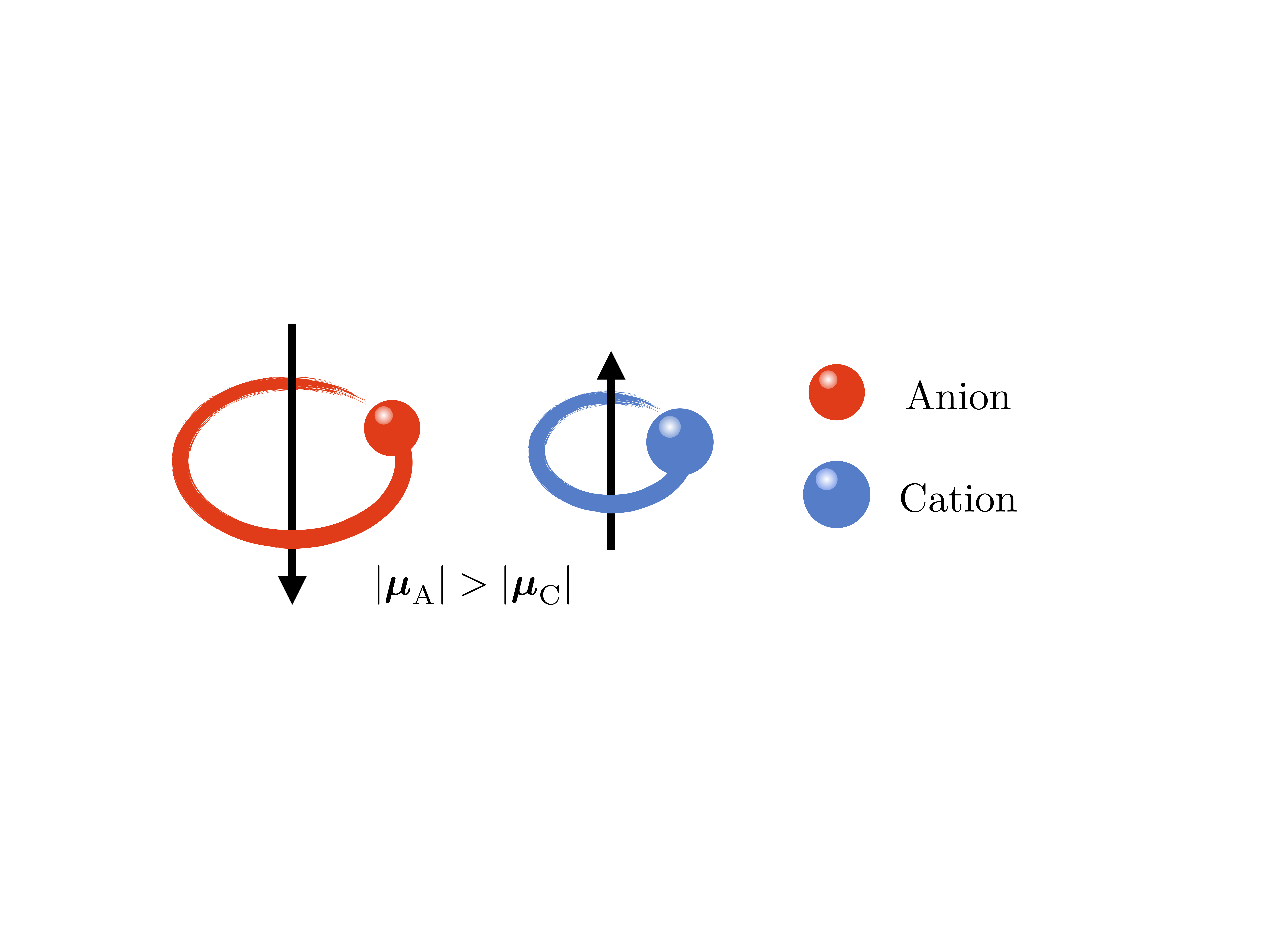}
\caption{
Schematic of the magnetic moments produced by the circular motion of ions along the eigenvectors of a circularly polarized optical phonon mode. Here, the lighter anion (drawn smaller) generates a larger magnetic moment $\mu_\text{A}$ than the heavier cation (drawn bigger) $\mu_\text{C}$, which in a binary material leads to the orbital magnetic moment of the phonon $\boldsymbol{\mu}_\text{ph} = \boldsymbol{\mu}_\text{A} + \boldsymbol{\mu}_\text{C}$ in the unit cell.
}
\label{fig:magmomschematic}
\end{figure}

When the atoms of a solid body move along the path of a circularly polarized vibration mode, they form closed loops and therefore carry angular momentum. In ionic materials, the circular motions of the ions induce orbital magnetic moments that are reminiscent of atomistic electromagnetic coils, see Fig.~\ref{fig:magmomschematic}, and that are linked to the angular momentum via the gyromagnetic ratio of the ions. As the ions have different masses, they move on different orbital radii and consequently produce magnetic moments that are unequal in size, leading to a net magnetic moment produced by the phonon mode. While the concept of rotational and vibrational angular momentum and magnetic moments has been discussed for over 80 years in molecules \cite{Wick1933,wick:1948,eshbach:1952,Huttner1970,anastassakis:1972,moss:1973,huttner:1978,Rebane:1983}, a rigorous microscopic \cite{ceresoli:2002,dzyaloshinskii:2009,dzyaloshinskii:2011,juraschek2:2017} and quantum mechanical \cite{riseborough:2010,zhang:2014,zhang:2015} treatment for solids has only been developed since the turn of the century.

In recent years, various physical phenomena have been attributed to the effect of phonon angular momentum, such as the phonon Hall and phonon spin Hall effects \cite{strohm:2005,sheng:2006,bliokh:2006}, a contribution to the Einstein-de Haas effect \cite{zhang:2014,Dornes2018,Mentink2018,Nakane2018,Streib2018} and to spin relaxation \cite{Garanin2015,Nakane2018}, the phonon a.c. Stark effect \cite{korenev:2016}, and the phonon Zeeman effect \cite{juraschek2:2017,juraschek2:2017}. Furthermore, terahertz sources are nowadays able to coherently excite phonons to yield large vibrational amplitudes, so that the effects of phonon angular momentum become visible also on a macroscopic level, for example by interaction with the magnetic or valley degrees of freedom of a material \cite{nova:2017,Shin2018}. In order to predict the strength of these effects accurately, it is necessary to calculate the governing physical parameters from first principles.

In this study, we calculate the size of the orbital magnetic moments of phonons in 35 different materials using a recently established microscopic formalism of the dynamical multiferroic effect \cite{juraschek2:2017}. The dynamical multiferroic effect describes in general the generation of magnetization from temporally varying electric polarization, $\mathbf{M} \propto \mathbf{P}\times\partial_t\mathbf{P}$ \cite{juraschek2:2017,Dunnett2018}, and $\mathbf{P}$ in our case corresponds to the dipole moment of an infrared (IR)-active phonon mode. We simulate the resonant excitation of coherent phonons with intense terahertz and mid-infrared radiation and calculate the size of the macroscopically induced orbital magnetic moments. We further compute the magnitude of the phonon Zeeman effect in an external magnetic field. As the orbital magnetic moment of a phonon is caused by the motion of the ions, we expect it to be of the order of the nuclear magneton, $\mu_\text{N}$, which is roughly three orders of magnitude smaller than the electronic magnetism, $\mu_\text{N}\approx 0.5\times10^{-3}\mu_\text{B}$. We therefore choose only nonmagnetic compounds, so that the magnetic signature of the material is caused entirely by the phonons. We choose sets of materials with each of the rocksalt, wurtzite, zincblende, and perovskite structures, to allow us to study trends within and across material classes. As a special case, we investigate a set of monolayer transition metal dichalcogenides, because they have recently been predicted and observed to host chiral phonons that are intrinsically circularly polarized \cite{zhang:2015,Zhu2018}.


\section{Theory of orbital magnetic moments of phonons}\label{sec:theory}


\subsection{Phonon angular momentum and phonon magnetic moment}

We begin by reviewing the mathematical expression for the angular momentum $\mathbf{L}$ and the magnetic moment $\mathbf{M}$ of a general elliptically polarized phonon mode, which is a superposition of two orthogonal linearly polarized phonon modes. $\mathbf{L}$ and $\mathbf{M}$, defined {\it per unit cell}, are given by
\begin{eqnarray}
\mathbf{L} & = & \mathbf{Q}\times\partial_t\mathbf{Q} \label{eq:phononangular},\\
\mathbf{M} & = & \gamma \mathbf{L}, \label{eq:phononmagmom}
\end{eqnarray}
where $\mathbf{Q}$ is the normal mode coordinate (or amplitude) vector in units of $\ang\sqrt{\text{amu}}$ with amu being the atomic mass unit, and $\gamma$ is the gyromagnetic ratio of the phonon \cite{ceresoli:2002,zhang:2014,juraschek2:2017}. The amplitude vector can be written without loss of generality as
\begin{eqnarray}\label{eq:amplitudevector}
\mathbf{Q}(t) = \left( \begin{array}{c}  Q_1(t)  \\
                             Q_2(t)  \\
                                      0
           \end{array}\right)
=
\left( \begin{array}{c}  Q_1 \sin(\Omega_1 t + \varphi)  \\
                             Q_2 \sin(\Omega_2 t)  \\
                                      0
           \end{array}
    \right),
\end{eqnarray}
where $\Omega_{1,2}$ are the eigenfrequencies of the two orthogonal phonon modes, and $\varphi$ is their relative phase shift. The angular momentum and magnetic moment accordingly reduce to $\mathbf{L}=L\hat{z}$ and $\mathbf{M}=M\hat{z}$, and we will in the following always refer to the $z$-components $L$ and $M$. The gyromagnetic ratio is given by
\begin{equation}\label{eq:gyromagneticratio}
\gamma=\sum_{i} \gamma_i \mathbf{q}_{i,1} \times \mathbf{q}_{i,2},
\end{equation}
where $\gamma_i=e\mathbf{Z}_i^\ast/(2\mathcal{M}_i)$ are the gyromagnetic ratios of the ions $i$, $\mathbf{Z}_i^\ast$ are the Born effective charge tensors, $\mathcal{M}_i$ are the masses, and $\mathbf{q}_{i,1/2}$ are the eigenvectors of the two superposed phonon modes \cite{juraschek2:2017}. $e$ denotes the elementary charge, and the index $i$ runs over all atoms in the unit cell. For circularly polarized phonon modes, $\Omega_1=\Omega_2\equiv\Omega_0$, $Q_1=Q_2\equiv Q$, $\varphi=\pi/2$, and the angular momentum in Eq.~(\ref{eq:phononangular}) simplifies to $\mathbf{L}=\mathbf{Q}\times\partial_t\mathbf{Q}=\Omega_0 Q^2\hat{z}\equiv L\hat{z}$.


\subsection{Oscillator model and ab-initio calculations}

In order to evaluate Eqs.~(\ref{eq:phononangular}) and (\ref{eq:phononmagmom}), we need to compute the gyromagnetic ratio $\gamma$ and the phonon amplitudes $Q_{1/2}$. The Born effective charge tensors $\mathbf{Z}^\ast_i$, and the phonon eigenvectors $\mathbf{q}_i$ and eigenfrequencies $\Omega_0$ are computed in this work using density functional theory. In order to obtain the phonon amplitude $Q$, we solve numerically the dynamical equation of motion
\begin{eqnarray}\label{eq:oscillatormodel}
\ddot{Q} + \kappa\dot{Q} + \partial_Q V(Q) = F(t).
\end{eqnarray}
For the effects that we investigate in this study, a minimal model consists of a harmonic potential $V(Q)=\Omega^2 Q^2/2$, zero damping $\kappa=0$, and a resonant mid-infrared driving force $F(t)=ZE(t)$. Here, $Z=\sum_i \mathbf{Z}^\ast_i \mathbf{q}_{i}/\sqrt{M_i}$ is the mode effective charge \cite{Gonze1997}, and we model the electric field of an ultrashort mid-infrared laser pulse as $E(t)=E_0 \text{exp}\{-(t-t_0)^2/[2(\tau/\sqrt{8\text{ln}2})^2]\}\text{cos}(\omega_0 t + \phi_\text{CEP})$, where $E_0$ is the peak electric field, $\tau$ is the full width at half maximum pulse duration, $\omega_0$ is 2$\pi$ times the center frequency, and $\phi_\text{CEP}$ is the carrier envelope phase \cite{Juraschek2018}. The amplitude and frequency of the excited phonons can be affected by nonlinear phonon couplings in the perovskites and by general anharmonicities in all investigated compounds. In Ref.~\cite{Juraschek2018} it was shown that the amplitudes of the IR-active phonon modes remain mostly unchanged in the presence of anharmonicities, and we therefore stay within the harmonic approximation.


\begin{figure}[b]
\centering
\includegraphics[scale=0.0765]{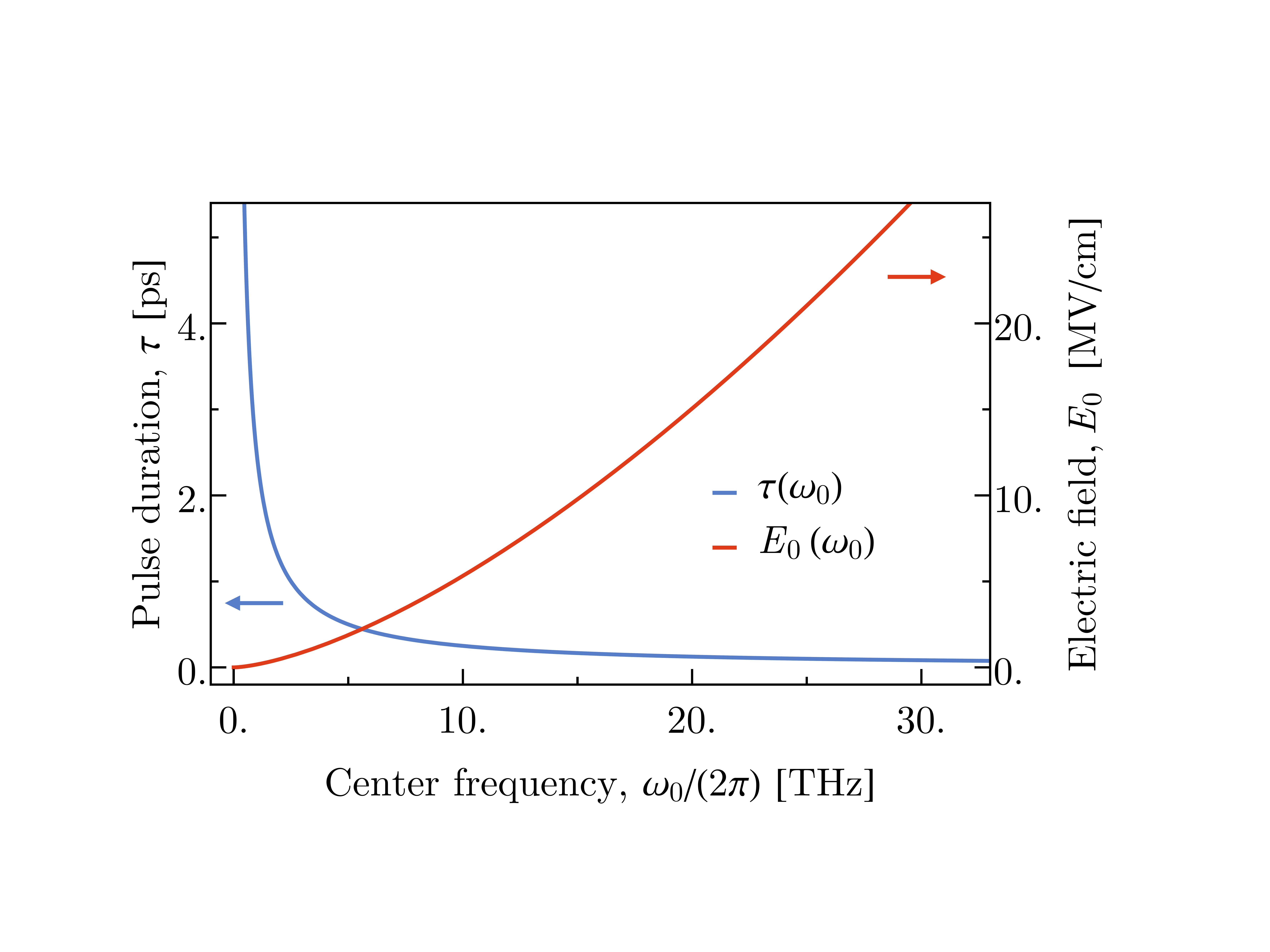}
\caption{
Dependence of the peak electric field $E_0$ and of the full width at half maximum pulse duration $\tau$ on the center frequency $\omega_0/(2\pi)$. We used a quadratic variation of $E_0$ and a variation of $\tau$ according to the ratio $\tau\omega_0 = 2.5\cdot2\pi$ in our calculations.
}
\label{fig:laserpulse}
\end{figure}

We calculated the phonon eigenfrequencies, eigenvectors, and the Born effective charge tensors from first-principles using the density functional theory formalism as implemented in the Vienna ab-initio simulation package (VASP) \cite{kresse:1996,kresse2:1996}, and the frozen-phonon method as implemented in the phonopy package \cite{phonopy}. We used the default VASP projector augmented wave (PAW) pseudopotentials for every considered atom and converged the Hellmann-Feynman forces to 50 $\mu$eV/\ang{}. For the rocksalt, wurtzite, zincblende, and perovskite structures we used plane-wave energy cut-offs of 700, 750, 750, and 700 eV, and 15$\times$15$\times$15, 12$\times$12$\times$8, 12$\times$12$\times$12, and 8$\times$8$\times$8 $k$-point gamma-centered Monkhorst-Pack meshes to sample the Brillouin zone, respectively \cite{Monkhorst/Pack:1976}. For the exchange-correlation functional, we chose the Perdew-Burke-Ernzerhof revised for solids (PBEsol) form of the generalized gradient approximation (GGA) \cite{csonka:2009}.


\begin{figure*}[t]
\centering
\includegraphics[scale=0.143]{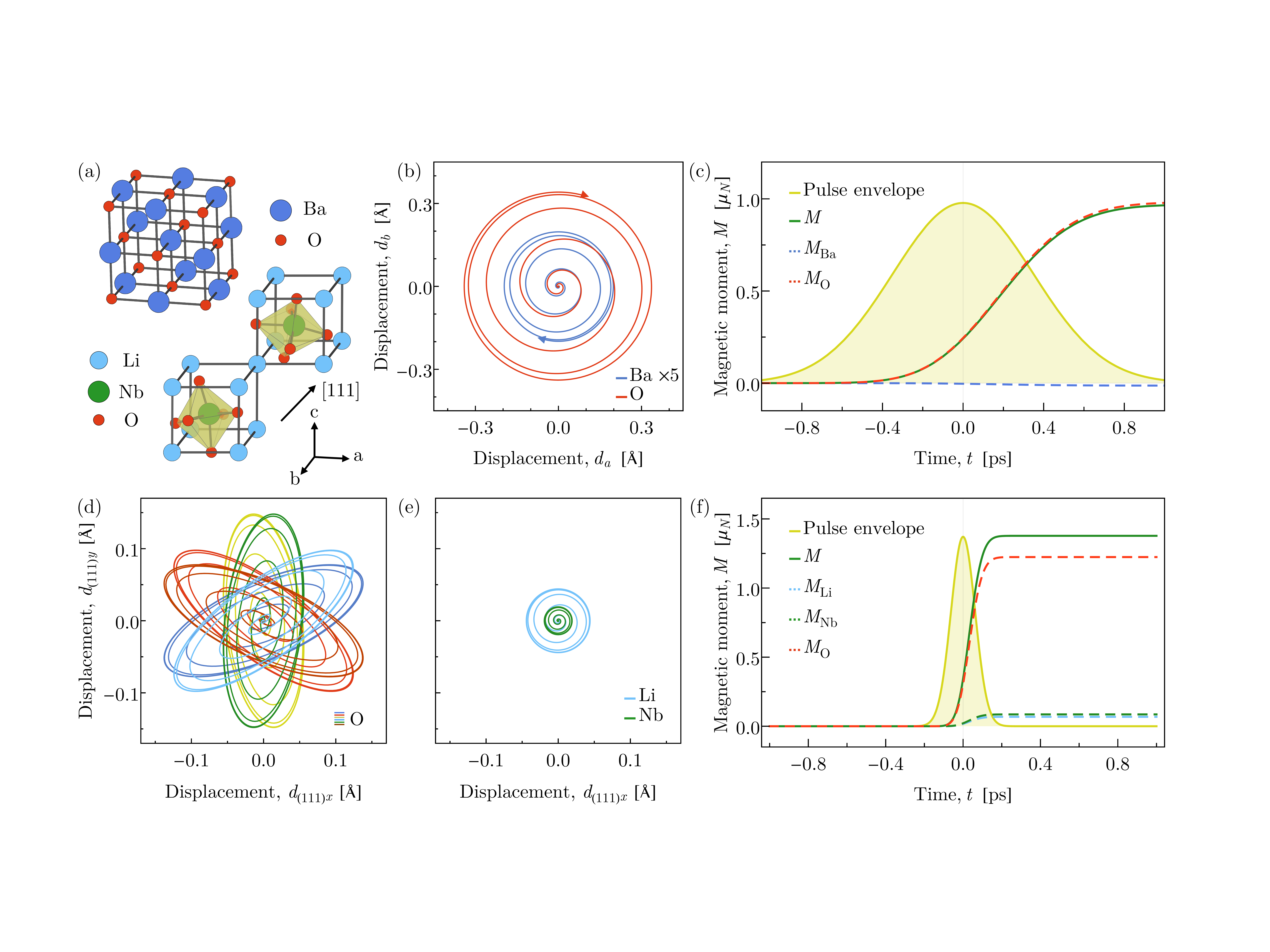}
\caption{
Visualization of the coherent phonon dynamics of the 3 THz mode in BaO and the 17 THz mode in LiNbO$_3$. All trajectories are shown for the time interval \{-1,1\} ps around the center of the laser pulse at $t_0=0$ and relative to their respective equilibrium positions (set to the origin of the plot). (a) Crystal structures of BaO and LiNbO$_3$. (b) Displacements of the barium and oxygen ions in the $ab$ plane of the BaO crystal. (c) Orbital magnetic moment $M$ per unit cell of the 3 THz mode and magnetic moments produced by each of the ions $M_\text{Ba}$ and $M_\text{O}$. The envelope of the pulse is shown schematically. (d) Displacements of the oxygen ions in the (111) plane of the LiNbO$_3$ crystal. (e) Displacements of the lithium and niobium ions in the (111) plane of the LiNbO$_3$ crystal. (c) Orbital magnetic moment $M$ per unit cell of the 17 THz mode and magnetic moments produced by each of the ions $M_\text{Li}$, $M_\text{Nb}$, and $M_\text{O}$. The envelope of the pulse is shown schematically.
}
\label{fig:phononplots}
\end{figure*}


\section{Results}


\begin{table*}[t]
\centering
\bgroup
\def\arraystretch{1.3}
\caption{
Binary compounds: IR-active phonon frequencies $\nu_0=\Omega_0/(2\pi)$ in THz, mode effective charges $Z$ in units of the elementary charge, root mean square displacements $d$ relative to the interatomic distances $d_0$ in percent, phonon population numbers $N$ per unit cell, phonon magnetons $\mu_\text{ph}$ and macroscopically induced orbital magnetic moments per unit cell $M=N\mu_\text{ph}$ in units of the nuclear magneton $\mu_\text{N}$, and phonon Zeeman splittings $\Delta\Omega/\Omega_0$ for an external magnetic field of $B=50~\text{T}$.
}
\begin{tabularx}{0.83\textwidth}{lccccccclccccccc}
\hline\hline
\multicolumn{1}{l}{Compound~~} & \multicolumn{1}{c}{$\nu_0$} & \multicolumn{1}{c}{$Z$} & \multicolumn{1}{c}{$d/d_0$} & \multicolumn{1}{c}{$N$} & \multicolumn{1}{c}{$\mu_\text{ph}$} & \multicolumn{1}{c}{$M$} & \multicolumn{1}{c}{$\Delta\Omega/\Omega_0$} &
\multicolumn{1}{l}{Compound~~} & \multicolumn{1}{c}{$\nu_0$} & \multicolumn{1}{c}{$Z$} & \multicolumn{1}{c}{$d/d_0$} & \multicolumn{1}{c}{$N$} & \multicolumn{1}{c}{$\mu_\text{ph}$} & \multicolumn{1}{c}{$M$} & \multicolumn{1}{c}{$\Delta\Omega/\Omega_0$} \\
\hline
\multicolumn{8}{l}{Rocksalt structure$^1$}                                         &   \multicolumn{8}{l}{Wurtzite structure$^3$}                                   \\
BaO   &   3.0   &   0.7   &   9   &   6   &   0.15   &   1.0   &   0.0002           &   BN   &   31.8   &   1.1   &   4   &   13   &   0.05   &   0.7   &  6$\times$10$^{-6}$   \\ 
CsF   &   3.6   &   0.3   &   3   &   1   &   0.06   &   0.1   &   0.00008         &   AlN   &   20.0   &   1.2   &   5   &   16   &   0.11   &   1.7   &   0.00002   \\ 
CsH   &   11.9   &   1.1   &   25   &   15   &   1.12   &   16.8   &   0.0005     &   GaN   &   17.1   &   1.1   &   6   &   15   &   0.17   &   2.5   &   0.00005   \\ 
LiI   &   4.8   &   0.5   &   7   &   3   &   0.18   &   0.5   &   0.0002          &   InN   &   14.7   &   1.2   &   6   &   16   &   0.2   &   3.2   &   0.00006   \\ 
MgO   &   11.7   &   0.6   &   4   &   5   &   0.04   &   0.2   &   0.00002        &   BeO   &   21.9   &   1.1   &   7   &   13   &   0.1   &   1.3   &   0.00002   \\ 
PbO$^2$   &   7.7   &   1.2   &   7   &   16   &   0.13   &   2.1   &   0.00008    &   CuH   &   31.0   &   0.7   &   12   &   6   &   0.49   &  2.7   &  0.00008   \\ 
PbS   &   2.2   &   0.8   &   8   &   8   &   0.12   &   1.0   &   0.0003          &   SiC   &   23.7   &   1.3   &   6   &   20   &   0.15   &   2.9   &   0.00003   \\
PbSe   &   1.6   &   0.6   &   4   &   5   &   0.04   &   0.2   &   0.0001         &   \multicolumn{8}{l}{Zincblende structure$^4$}                                     \\
PbTe   &   1.4   &   0.7   &   3   &   5   &   0.02   &   0.1   &   0.00006        &   BeS   &   17.3   &   0.6   &   5   &   4   &   0.13   &   0.5   &   0.00004   \\       
SnTe   &   1.0   &   1.0   &   5   &   13   &   0.004   &   0.1   &   0.00002      &   BeSe   &   15.3   &   0.5   &   5   &   3   &   0.15   &   0.5   &   0.00005   \\ 
       &         &          &       &           &           &           &         &   BeTe   &   14.1   &   0.4   &   4   &   2   &   0.14   &   0.3   &   0.00005   \\ 
       &         &          &       &           &           &           &         &   GaAs   &   7.9   &   0.4   &   1   &   2   &   0.002   &   0.004   &    1$\times$10$^{-6}$   \\ 
\multicolumn{16}{l}{$^1$cubic (Fm$\bar{3}$m) $^2$tetragonal (P4/nmm) $^3$hexagonal (P6$_3$cm) $^4$cubic (F$\bar{4}$3m)} \\
\hline\hline
\end{tabularx}
\label{tab:binaries}
\egroup
\end{table*}


\subsection{Orbital magnetic moments of coherent phonons and single phonon quanta}



We turn to the numerical results obtained using Eq.~(\ref{eq:phononmagmom}). In order to investigate comparable excitation strengths in the different materials, we scale the fluence of the laser pulse linearly with the energy $\hbar\Omega_0$ of the resonantly driven phonons. We keep the number of cycles of the simulated laser pulse constant by fixing the ratio $\tau/\omega_0 = 2.5 \cdot 2\pi$, and we scale the peak electric field quadratically with the laser frequency, taking $E_0 = 15~\text{MV/cm}$ at $\omega_0/(2\pi)=20~\text{THz}$ as our reference point, see Fig.~\ref{fig:laserpulse}. We found this to be an appropriate way to scale the electric field, as fixing the fluence instead leads to an overproportional excitation of low-frequency phonon modes.

In Fig.~\ref{fig:phononplots} we show the time-dependent responses of the atoms to the coherent excitation by the laser pulse; for the 3 THz mode in barium oxide (BaO) (upper row), and for the 17 THz mode in lithium niobate (LiNbO$_3$) (lower row). Their crystal structures are shown in Fig.~\ref{fig:phononplots}a. We plot the trajectories of the ions for the time interval \{-1,1\} ps around the center of the laser pulse at $t_0$=0. The displacement of the ions is shown for the $ab$ plane of the BaO crystal in Fig.~\ref{fig:phononplots}b and for the (111) plane of the LiNbO$_3$ crystal in Figs.~\ref{fig:phononplots}d and e. The displacements are shown with respect to the equilibrium positions of each ion, which are set to the origin of the plots. We further show the time evolution of the orbital magnetic moment $M$ of the circularly polarized phonon mode, and of the constituent magnetic moments $M_i$ generated by the circular motion of each ion $i$ in Figs.~\ref{fig:phononplots}c and f. The envelope of the pulse is shown schematically. 

In BaO, the radius of the circular motion of barium is small compared to that of oxygen, and the corresponding magnetic moment $M_\text{Ba}$ (scaling quadratically with the radius) is negligible. $M$ is therefore almost entirely generated by the motion of the oxygen ions. Here (and in the binary compounds in general), the magnetic moments of the anions and cations have opposite signs. In LiNbO$_3$, the mass differences between the ions are not as big as in BaO, and the corresponding magnetic moments of the cations $M_\text{Li}$ and $M_\text{Nb}$ contribute notably to the orbital magnetic moment of the phonon mode. Here, the eigenvector of the circularly polarized phonon mode is such that the magnetic moment of the cations and anions are cooperative.

We now formulate the equations of Sec.~\ref{sec:theory} in terms of single phonon quanta. The harmonic vibrational energy per unit cell of the two superposed phonon modes is given by $V_\text{vib}=\Omega_1^2 Q_1^2/2 + \Omega_2^2 Q_2^2/2 \equiv \Omega^2_0 Q^2$ and the energy of a single phonon quantum by $\hbar \Omega_0$. We obtain the phonon population number $N$ per unit cell as 
\begin{equation}\label{eq:populationnumber}
N=\frac{V_\text{vib}}{\hbar\Omega_0}=\frac{\Omega_0}{\hbar} Q^2,
\end{equation}
and subsequently rewrite $L$ and $M$ in terms of $N$ as
\begin{eqnarray}
L & = & N \hbar \equiv N l_\text{ph}, \label{eq:Nphononangular} \\
M & = & \gamma N \hbar \equiv N \mu_\text{ph}. \label{eq:Nphononmagmom}
\end{eqnarray}
Here, the quantized angular momentum of a circularly polarized phonon is equal to the reduced Planck constant $l_\text{ph}=\hbar$ \cite{zhang:2014,zhang:2015}, and its quantized magnetic moment is given by $\mu_\text{ph}=\gamma l_\text{ph}=\gamma\hbar$, which we will refer to in the following as the ``phonon magneton''. Note that previous publications sometimes refer to the angular momentum of phonons as ``phonon spin'' \cite{Garanin2015,zhang:2015,Holanda2018}. Since the phonon is a quasiparticle and the angular momentum arises from circular (orbital) motions of the atoms, we find it more appropriate to use the term ``intrinsic orbital angular momentum'' \footnote{``Extrinsic orbital angular momentum'' would correspond to a macroscopic rotation or elastic deformation of the crystal}.


\begin{table*}[t]
\centering
\bgroup
\def\arraystretch{1.3}
\caption{
Perovskites in their low-temperature structures: IR-active phonon frequencies $\nu_0=\Omega_0/(2\pi)$ in THz, mode effective charges $Z$ in units of the elementary charge, root mean square displacements $d$ relative to the interatomic distances $d_0$ in percent, phonon population numbers $N$ per unit cell, phonon magnetons $\mu_\text{ph}$ and macroscopically induced orbital magnetic moments per unit cell $M=N\mu_\text{ph}$ in units of the nuclear magneton $\mu_\text{N}$, and phonon Zeeman splittings $\Delta\Omega/\Omega_0$ for an external magnetic field of $B=50~\text{T}$. We display a selection of phonons with the largest values of $\nu_0$, $N$, $\mu_\text{ph}$, and $M$ for each compound.
}
\begin{tabularx}{0.815\textwidth}{lccccccclccccccc}
\hline\hline
\multicolumn{1}{l}{Compound~~} & \multicolumn{1}{c}{$\nu_0$} & \multicolumn{1}{c}{$Z$} & \multicolumn{1}{c}{$d/d_0$} & \multicolumn{1}{c}{$N$} & \multicolumn{1}{c}{$\mu_\text{ph}$} & \multicolumn{1}{c}{$M$} & \multicolumn{1}{c}{$\Delta\Omega/\Omega_0$} &
\multicolumn{1}{l}{Compound~~} & \multicolumn{1}{c}{$\nu_0$} & \multicolumn{1}{c}{$Z$} & \multicolumn{1}{c}{$d/d_0$} & \multicolumn{1}{c}{$N$} & \multicolumn{1}{c}{$\mu_\text{ph}$} & \multicolumn{1}{c}{$M$} & \multicolumn{1}{c}{$\Delta\Omega/\Omega_0$} \\
\hline
BaHfO$_3$$^1$  &  15.7  &  0.8  &  5  &  8  &  0.04  &  0.3  &  0.00001  &   LiTaO$_3$$^2$  &  17.4  &  1.5  &  5  &  25  &  0.04  &  1.0  &  0.00001  \\
        &  5.9  &  1.1  &  7  &  14  &  0.12  &  1.7  &  0.0001  &            &  10.9  &  0.9  &  8  &  10  &  0.13  &  1.3  &  0.00006  \\
BaZrO$_3$$^1$  &  15.0  &  0.9  &  6  &  10  &  0.04  &  0.4  &  0.00001  &          &  4.2  &  1.0  &  4  &  11  &  0.07  &  0.8  &  0.00008  \\
        &  5.8  &  1.0  &  5  &  12  &  0.04  &  0.4  &  0.00003  &   BaTiO$_3$$^3$  &  14.1  &  0.8  &  3  &  8  &  0.03  &  0.3  &  0.00001  \\
        &  3.1  &  0.7  &  5  &  6  &  0.07  &  0.4  &  0.0001  &             &  8.9  &  0.01  &  --  &  --  &  0.14  &  --  &  0.00007  \\
KTaO$_3$$^1$   &  15.8  &  1.0  &  6  &  12  &  0.003  &  0.04  &  9$\times$10$^{-7}$  &           &  6.5  &  2.2  &  9  &  58  &  0.02  &  1.0  &  0.00001  \\
        &  2.3  &  1.4  &  14  &  24  &  0.12  &  3.0  &  0.0003  &   KNbO$_3$$^3$   &  15.2  &  1.7  &  7  &  35  &  0.01  &  0.5  &  5$\times$10$^{-6}$  \\
BiAlO$_3$$^2$  &  18.5  &  0.6  &  2  &  4  &  0.08  &  0.3  &  0.00002  &           &  8.1  &  0.02  &  --  &  --  &  0.18  &  --  &  0.0001  \\
        &  12.8  &  0.1  &   --   &   --    &  0.14  &   --    &  0.00005  &            &  6.0  &  2.2  &  10  &  55  &  0.13  &  7.0  &  0.0001  \\ 
        &  11.5  &  1.4  &  6  &  22  &  0.05  &  1.1  &  0.00002  &  PbTiO$_3$$^4$  &  14.9  &  0.8  &  6  &  8  &  0.09  &  0.7  &  0.00003  \\
        &  3.9  &  1.1  &  6  &  14  &  0.05  &  0.7  &  0.00006  &           &  8.1  &  0.1  &  1  &  0.2  &  0.19  &  0.05  &  0.0001  \\
CsPbF$_3$$^2$  &  9.1  &  0.04  &   --    &    --   &  0.001  &       &  3$\times$10$^{-7}$  &              &  2.6  &  0.8  &  7  &  7  &  0.08  &  0.6  &  0.0001 \\
        &  4.9  &  0.6  &  3  &  4  &  0.04  &  0.2  &  0.00004  &    SrTiO$_3$$^5$  &  15.7  &  1.1  &  4  &  15  &  0.03  &  0.4  &  9$\times$10$^{-6}$  \\
        &  1.5  &  0.02  &   --    &   --    &  0.04  &   --    &  0.0001  &               &  7.3  &  0.04  &  --  &  --  &  0.18  &  --  &  0.0001  \\
LiNbO$_3$$^2$  &  17.0  &  1.6  &  6  &  32  &  0.04  &  1.4  &  0.00001  &          &  1.7  &  2.5  &  21  &  74  &  0.1  &  7.2  &  0.0003  \\
        &  10.6  &  0.8  &  7  &  7  &  0.13  &  0.9  &  0.00006  & & & & & & & & \\
        &  4.3  &  1.0  &  6  &  12  &  0.1  &  1.2  &  0.0001    & & & & & & & & \\
\multicolumn{16}{l}{
$^1$cubic (Pm$\bar{3}$m)
$^2$rhombohedral (R3c)
$^3$rhombohedral (R3m)
} \\
\multicolumn{16}{l}{
$^4$tetragonal (P4mm)
$^5$tetragonal (I4/mcm)
} \\
\hline\hline
\end{tabularx}
\label{tab:perovskites}
\egroup
\end{table*}

In Tables~\ref{tab:binaries} and \ref{tab:perovskites} we show the results of our calculations for the binary and perovskite compounds, respectively. We show the calculated phonon eigenfrequencies $\nu_0=\Omega_0/(2\pi)$ and mode effective charges $Z$, the phonon magnetons $\mu_\text{ph}=\gamma\hbar$ obtained using Eq.~(\ref{eq:gyromagneticratio}), the phonon population numbers $N$ per unit cell obtained from Eqs.~(\ref{eq:oscillatormodel}) and (\ref{eq:populationnumber}), and the macroscopically induced orbital magnetic moment per unit cell $M$ obtained from Eq.~(\ref{eq:Nphononmagmom}). To check that the vibrational response to the pulsed excitation is in a meaningful range, we evaluate the Lindemann criterion for the induced atomic displacements. According to the Lindemann criterion, melting occurs when the root mean square displacement $d$ of the atoms reaches around 10-20\%{} of the interatomic distance $d_0$ \cite{Lindemann1910,sokolowski-tinten:2003}. We therefore evaluate $d$ for the ion $i$ with the largest root mean square displacement $\mathbf{d}_i=\mathbf{q}_i Q/\sqrt{2\mathcal{M}_i}$ as $d=\text{max}_i | \mathbf{d}_i |$, and we include the ratio $d/d_0$ in the Tables~\ref{tab:binaries} and \ref{tab:perovskites}.

The phonon magneton is enhanced by high Born effective charges and big mass differences between the ions, as is apparent from Eq.~(\ref{eq:gyromagneticratio}). As a result, we found the largest phonon magnetons of $\mu_\text{ph}=1.2$ and $0.49~\mu_\text{N}$ in the hydride compounds CsH and CuH, because of the large mass difference due to the light H atom. Typical values for the other binary and perovskite compounds range between $\mu_\text{ph}\sim0.1$ and $0.2~\mu_\text{N}$.

When looking at the experimental feasibility of inducing large values of $M$, two factors in addition to the phonon magneton have to be taken into account: the excitability of the phonon mode and the limitation due to the Lindemann criterion. As we scale the fluence of the laser pulse with the energy of the phonons, the mode effective charge $Z$ is the main factor in determining the excitability of the phonon mode. It determines the amplitude of the driven phonon mode according to Eq.~(\ref{eq:oscillatormodel}), and consequently the population number $N$ per unit cell according to Eq.~(\ref{eq:populationnumber}). For example, the exceptionally large mode effective charges of $Z>2e$ of the 6~THz modes in BaTiO$_3$ and KNbO$_3$ and the soft 1.7~THz mode in SrTiO$_3$ lead to high population numbers of $>50$ phonons per unit cell under the pulsed excitation. This excitability is however limited by the Lindemann criterion, which predicts instability of the lattice due to melting for values of $d/d_0>10\%$. In most cases we obtain phonon amplitudes within the stability limit, however the previous example shows that while the phonon population in KNbO$_3$ leads to $d/d_0=10\%$, the corresponding excitation of the soft mode in SrTiO$_3$ with $d/d_0=21\%$ would likely destroy the sample, and the peak electric field $E_0$ has to be scaled down accordingly. The most convenient materials for achieving large $M$ are therefore PbO, AlN, GaN, InN, and SiC for the binary compounds (CsH can be omitted for practical reasons, due to its instability under air), in which values of $M\sim1.7$ to $3.2~\mu_\text{N}$ should be achievable. For the perovskites, we calculated values of $M=1.7$ and $3.0~\mu_\text{N}$ in BaHfO$_3$ and KTaO$_3$, and up to $M\sim7~\mu_\text{N}$ for KNbO$_3$ and SrTiO$_3$, due to the presence of phonon modes with both high excitability and large phonon magnetons.


\subsection{Phonon Zeeman effect}

The orbital magnetic moment of the phonon interacts with an external magnetic field $\mathbf{B}=B\hat{z}$ via Zeeman coupling of the form $\mathbf{M}\cdot\mathbf{B}$. This leads to a Zeeman splitting of the phonon frequencies of the right (+) and left (--) handed circular polarization of the phonon that is linear in the external magnetic field \cite{ceresoli:2002,juraschek2:2017}:
\begin{eqnarray}
\Omega_\pm = \Omega_0 \pm \gamma B. \label{eq:Zeemansplitting}
\end{eqnarray}
The relative splitting of the phonon frequency is hence given by
\begin{eqnarray}
\frac{\Delta\Omega}{\Omega_0} = \frac{2\gamma B}{\Omega_0} = \frac{2\mu_\text{ph}B}{\hbar\Omega_0}. \label{eq:relativesplitting}
\end{eqnarray}
The effect is independent of the phonon population number. We therefore have to compute only $\mu_\text{ph}$ in order to calculate the magnitude of the effect.

We show the calculated relative splittings of the binary compounds in Table~\ref{tab:binaries}, and of the perovskites in Table~\ref{tab:perovskites}, in an external magnetic field of $B=50~\text{T}$. The splitting is highest for low-frequency phonons with high $\mu_\text{ph}$. The largest splittings of $\Delta\Omega/\Omega_0>10^{-4}$ occur for BaO, CsH, LiI, PbS, PbSe for the binary compounds, and are comparably large for most of the perovskites, in which low-frequency phonons are generally present.


\subsection{Orbital magnetic moments of chiral phonons}

We now discuss the orbital magnetic moments produced by chiral phonons, which have recently been proposed and observed in the valleys of the monolayer transition metal dichalcogenide (TMD) WSe$_2$ and other graphene-derived materials \cite{zhang:2015,Zhu2018,Gao2018,Xu2018}. The motion of the ions along the eigenvectors of a chiral phonon mode is intrinsically circular and cannot be constructed by a superposition of linearly polarized phonon modes. Chiral phonons are generated through the decay of an exciton state, in which a hole scatters between the K/K' points (valleys) of the Brillouin zone under emission of a chiral phonon. Because they occur at nonzero wave vectors, they cannot be coherently excited via IR-absorption, which is only possible for modes at the center of the Brillouin zone.

The calculation of $\mu_\text{ph}=\gamma\hbar$ using Eq.~(\ref{eq:gyromagneticratio}) has to be adjusted: the eigenvectors $\mathbf{q}_{i,1}$ and $\mathbf{q}_{i,2}$ no longer correspond to the eigenvectors of two superposed phonon modes, but to the real and imaginary parts $\text{Re}[\mathbf{q}_{i,\text{chiral}}]$ and $\text{Im}[\mathbf{q}_{i,\text{chiral}}]$ of the eigenvector that defines the circular motions of the ions in the chiral phonon modes. Here, the transition metal stands still, while the entire vibrational motion is made by the respective chalcogen atoms. We calculated the phonon magnetons $\mu_\text{ph}$ for the series of WS$_2$, WSe$_2$, and WTe$_2$, and we list the values in Table~\ref{tab:TMDs}. The phonon magnetons with a maximum value of $M=0.0007~\mu_\text{N}$ for WS$_2$ are small compared to the other material classes, because the chalcogen atoms that make the circular motion possess only small Born effective charges of $\mathbf{Z}^\ast_i \sim 0.3$ to $1.3e$.


\begin{table}[t]
\centering
\bgroup
\def\arraystretch{1.5}
\caption{
Monolayer transition metal dichalcogenides (TMDs) in their hexagonal (P$\bar{6}$m2) structure: Chiral phonon frequencies $\nu_0$ in THz and phonon magnetons $\mu_\text{ph}$ in units of the nuclear magneton $\mu_\text{N}$ at the K/K' points of the Brillouin zone.
}
\begin{tabularx}{0.43\textwidth}{lcclcclcc}
\hline\hline
\multicolumn{1}{l}{TMD} & \multicolumn{1}{c}{$\nu_0$} & \multicolumn{1}{c}{$\mu_\text{ph}$} &
\multicolumn{1}{l}{TMD} & \multicolumn{1}{c}{$\nu_0$} & \multicolumn{1}{c}{$\mu_\text{ph}$} &
\multicolumn{1}{l}{TMD} & \multicolumn{1}{c}{$\nu_0$} & \multicolumn{1}{c}{$\mu_\text{ph}$} \\
\hline
WS$_2$ & 6.7 & 0.0007 & WSe$_2$ & 6.4  & 0.0003 & WTe$_2$ & 6.0  & 0.0002  \\
\hline\hline
\end{tabularx}
\label{tab:TMDs}
\egroup
\end{table}


\begin{figure*}[t]
\centering
\includegraphics[scale=0.13]{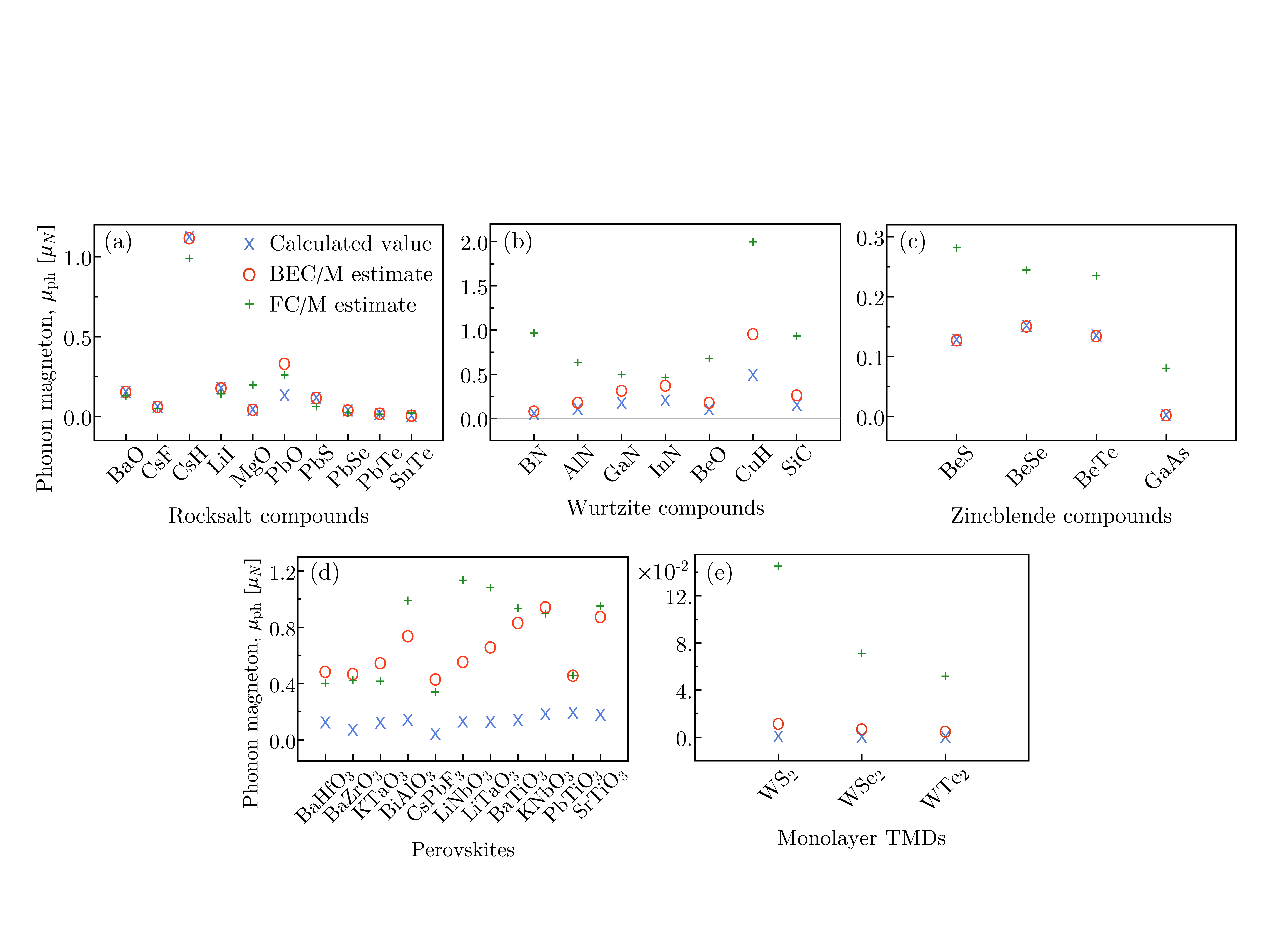}
\caption{
Calculated values of the phonon magneton $\mu_\text{ph}$ and estimated values using the Born effective charge (BEC) and formal charge (FC) to mass ratio. (a) Rocksalt, (b) wurtzite, (c) zincblende compounds, (d) perovskites, and (e) monolayer transition metal dichalcogenides (TMDs). For the perovskites in (d), we show the phonon modes displaying the highest $\mu_\text{ph}$ for each compound.
}
\label{fig:estimate}
\end{figure*}


\subsection{A simple estimate of the phonon magneton}

In this last results section, we investigate whether it is possible to derive a simple estimate of the magnitude of the phonon magneton without detailed knowledge about the phonon eigenvectors. If we write the phonon magneton as given by Eq.~(\ref{eq:gyromagneticratio}) in units of the nuclear magneton $\mu_\text{N}=e\hbar/(2\mathcal{M}_\text{P})$, with the proton mass $\mathcal{M}_\text{P}\approx 1~\text{amu}$, we remain with $\mu_\text{ph}=\sum_{i} \mathbf{Z}_i^\ast/\mathcal{M}_i (\mathbf{q}_{i,1} \times \mathbf{q}_{i,2})$. Therefore, poor-man's estimates, which we denote $S$ and in which we ignore the contribution of the phonon eigenvectors, can be written as
\begin{eqnarray}
S_\text{BEC/M} & = & \sum\limits_i \frac{\text{Tr}[\mathbf{Z}_i^\ast]}{3\mathcal{M}_i}, \label{eq:BECestimate}\\
S_\text{FC/M} & = & \sum\limits_i \frac{Z_i^\text{(f)}}{\mathcal{M}_i}. \label{eq:FCestimate}
\end{eqnarray}
Both estimates consist of a sum of the charge-to-mass ratio of the ions. In $S_\text{BEC/M}$, we take the average trace Tr[$\mathbf{Z}_i^\ast$]/3 of the Born effective charge (BEC) tensor, for which literature values are often available and density functional theory calculations are inexpensive. In $S_\text{FC/M}$, we simply use the formal charges (FC) $Z_i^\text{(f)}$, where the entire information can be extracted from the periodic table of the elements.

In Fig.~\ref{fig:estimate} we show a comparison of the estimates $S_\text{BEC/M}$ and $S_\text{FC/M}$ to the calculated values of $\mu_\text{ph}$ for each of the materials classes rocksalt, wurtzite, zincblende, perovskite, and monolayer transition metal dichalcogenides. For the materials with diatomic unit cells (rocksalt and zincblende structures), the estimate from the sum over the Born effective charge to mass ratio $S_\text{BEC/M}$ is in excellent agreement with the calculated values, see Figs.~\ref{fig:estimate}a and c. For the materials with 4-atom unit cells (wurtzite structure and PbO), $S_\text{BEC/M}$ deviates from the calculated value by a significant margin - it does however capture the relative trend within the materials class and can be used to estimate an upper boundary for $\mu_\text{ph}$, see Fig.~\ref{fig:estimate}b. For the perovskites with unit cells consisting of 5 or 10 atoms, $S_\text{BEC/M}$ is no longer a good predictor, see Fig.~\ref{fig:estimate}d, because the vector product of the phonon eigenvectors $\mathbf{q}_{i,1} \times \mathbf{q}_{i,2}$ is different for each of the sets of degenerate IR-active phonon modes. Furthermore, the shape of the Born effective charge tensors with their inhomogeneous diagonal and nonzero off-diagonal terms can no longer be accounted for. A similar analysis can be applied to the monolayer transition metal dichalcogenides with 3-atom unit cells, in which the trend across the series is captured, but $\mu_\text{ph}$ is strongly overestimated due to the large inhomogeneity of the diagonal Born effective charges parallel and normal to the two-dimensional surface. Finally, the sum over the formal charge to mass ratio $S_\text{FC/M}$ predicts neither the magnitude of $\mu_\text{ph}$, nor the trends within materials classes, which emphasizes the importance of using the Born effective charge formalism to include the electronic rehybridization contribution to the electric dipole moment of the IR-active phonon modes.


\section{Conclusion}

In summary, we find the quantized orbital moments of circularly polarized phonons, which we call phonon magnetons, to be in the order of $10^{-4}\mu_\text{B}$ per unit cell, and the macroscopically induced orbital magnetic moments of coherent phonons generated from pulsed mid-IR excitation to reach the order of $10^{-3}\mu_\text{B}$ per unit cell. The phonon Zeeman splittings in an external magnetic field of 50 T reach the order of 10$^{-4}$ of the phonon frequency. We expect that the orbital magnetic moments of phonons should be observable with modern experimental techniques, for example indirectly via Faraday rotation measurements that are able to detect small changes in electronic magnetic order \cite{nova:2017}, or directly via nitrogen-vacancy center magnetometry \cite{Degen2008,Maze2008,Taylor2008,Casola2018}. We hope that our analysis of the information generated from our database will stimulate and guide future experimental studies in the field of optical phononics.


\begin{acknowledgments}
We thank Youichi Yanase (Kyoto University) for useful discussions. This work was supported by the ETH Z\"{u}rich and by the JSPS KAKENHI Grant No. JP15K21732 (J-Physics). Calculations were performed at the Swiss National Supercomputing Centre (CSCS) supported by the project IDs s624 and p504.
\end{acknowledgments}



%

\end{document}